\documentclass[12pt]{article}
\usepackage{amssymb}
\def\be{\begin{equation}}
\def\ee{\end{equation}}
\def\bea{\begin{eqnarray}}
\def\eea{\end{eqnarray}}

\def\halfj{{\textstyle\frac{j}{2}}}
\def\halfhalf{{\textstyle\frac{1}{4}}}

\def\pl#1{{\sl Phys.~Lett.~\bf B#1}}
\def\pr#1{{\sl Phys.~Rev.~\bf D#1}}
\def\prl#1{{\sl Phys.~Rev. Lett.~\bf #1}}

\def\cqg#1{{\sl Class.~Quant.~Grav.~\bf #1}}

\begin{document}
\hfill UTHET-04-0901

\vspace{-0.2cm}

\begin{center}
\Large
{\bf Perturbative calculation of quasi-normal modes\footnote{Presented at PASCOS 2004 / Nath Fest.}}
\normalsize

\vspace{.5cm}

{\sc George Siopsis}\footnote{Research supported in part by the US Department of Energy under grant
DE-FG05-91ER40627.}\\ {\em Department of Physics
and Astronomy,} \\
{\em The University of Tennessee, Knoxville,} \\
{\em TN 37996 - 1200, USA.} \\
E-mail: {\tt siopsis@tennessee.edu}

\end{center}

\vspace{0.8cm}
\large
\centerline{\bf Abstract}
\normalsize
\vspace{.5cm}

I discuss a systematic method of analytically calculating the asymptotic form of quasi-normal frequencies. In the case of a four-dimensional Schwarzschild black hole, I expand around the zeroth-order approximation to the wave equation proposed by Motl and Neitzke. In the case of a five-dimensional AdS black hole, I discuss a perturbative solution of the Heun equation. The analytical results are in agreement with the results from numerical analysis.

\newpage

\section{Introduction}

This report summarizes recent work I did largely in collaboration with S.~Musiri~\cite{bibus,bibus2,bibus3}.

Quasi-normal modes (QNMs) describe small perturbations of a black hole.
They are obtained by solving
a wave equation for small fluctuations subject to the conditions that the
flux be
ingoing at the horizon and
outgoing at asymptotic infinity.
In general, one obtains a discrete spectrum of complex frequencies.
The imaginary part determines the decay time of the small fluctuations
($\Im \omega = \frac{1}{\tau}$).

\section{Schwarzschild black holes}


For a Schwarzschild black hole,
\be \omega_n/T_H = (2n+1)\pi i + \ln 3 \ee
This was first derived numerically~\cite{bibn1,bibn2,bibn3,bibn4,bibn5}
and subsequently confirmed analytically~\cite{bibx3}.
$\Im\omega_n$ is large and
easy to understand because the spacing of frequencies is $2\pi i T_H$,
same as the spacing of poles of a thermal Green function.
On the other hand,
$\Re\omega_n$ is small.
Its analytical value was conjectured by Hod~\cite{bibhod}
and is related to the Barbero-Immirzi parameter. It is intriguing
from the loop quantum gravity point of view suggesting that the gauge group
should be $SO(3)$ rather than $SU(2)$. Thus the study of QNMs
may lead to a deeper understanding of black holes and quantum gravity.

The analytical derivation of the asymptotic form of QNMs~\cite{bibx3} offered a new surprise as it heavily
relied on the black hole singularity.
It is intriguing that the {\it unobservable}
region beyond the horizon influences the behavior of physical quantities.
In the Schwarzschild background metric the wave equation for a spin-$j$ perturbation
of frequency $\omega$ is
\be
-\frac{d^2\Psi}{d z^2} + \frac{1}{\omega^2}\ V[r(z)] \Psi  =\Psi \ , \ V(r) = \left( 1 - {\textstyle\frac{r_0}{r}} \right) \big( {\textstyle\frac{\ell (\ell +1)}{r^2}} + \textstyle{\frac{(1-j^2)r_0}{r^3}} \big)\ee
where $z = \omega (r+ r_0 \ln (r/r_0 -1) -i\pi r_0)$ and $r_0$ is the radius of the horizon.
For QNMs we demand
(assuming $\Re \omega > 0$)
$e^{iz} \Psi\sim 1$ as $z\to +\infty$
and near the horizon ($z\to -\infty$),
$e^{iz} \Psi \sim e^{2iz}$.
The latter boundary condition is implemented by demanding that the monodromy
around the singular point $r=r_0$ in the complex $r$-plane be
$\mathcal{M} (r_0) = e^{-4\pi\omega r_0}$
along a contour running counterclockwise.
We may deform the contour so that the only contribution to the monodromy
comes from the black hole singularity ($r=0$)~\cite{bibx3}.
The potential can be written as a series in $\sqrt{z}$,
which is also a formal expansion in $1/\sqrt\omega$.
We may then solve the wave equation perturbatively.
To zeroth order, we obtain
\be
\Psi_0'' + \big\{ {\textstyle\frac{1-j^2}{4z^2}} + 1 \big\} \Psi_0 = 0 \ee
whose solutions can be written in terms of Bessel functions.
They lead to a zeroth-order monodromy~\cite{bibx3}
$\mathcal{M} (r_0) = - (1+ 2\cos (\pi j))$
yielding a discrete spectrum of complex frequencies (QNMs)
\be\label{eqqnm0} \omega_n/T_H = (2n+1)\pi i + \ln (1+2\cos(\pi j) ) + o(1/\sqrt n) \ee
Expanding the wavefunction,
$\Psi = \Psi_0 + \frac{1}{\sqrt{-\omega r_0}}\, \Psi_1 + o(1/\omega)$,
the first-order correction obeys
\be\label{eqfo}
\Psi_1'' + \big\{ {\textstyle\frac{1-j^2}{4z^2}} + 1 \big\} \Psi_1  = \delta V \Psi_0 \ , \ \delta V = -{\textstyle\frac{3\ell (\ell+1) +1- j^2 }{6\sqrt 2}}\, z^{-3/2}\ee
By solving eq.~(\ref{eqfo}), one obtains an $o(1/\sqrt n)$ correction to the
monodromy which yields the
QNM frequencies~\cite{bibus}
$$\omega_n/T_H = (2n+1)\pi i + \ln (1+2\cos(\pi j) )
+ {\textstyle\frac{e^{\pi ji/2}}{\sqrt{n+1/2}}} \mathcal{A}
+ o(1/n) \ \ , $$
\be  \mathcal{A}
= (1-i) \ {\textstyle\frac{3\ell(\ell +1) +1-j^2}{24\sqrt 2 \pi^{3/2}}
\ \frac{\sin (2\pi j)}{\sin (3\pi j/2)}}
\ \Gamma^2 (\halfhalf)\ \Gamma(\halfhalf + \halfj)
\ \Gamma (\halfhalf - \halfj)\ee
This analytical result is in agreement with numerical results for scalar ($j=0$)
and gravitational waves ($j=2$).~\cite{bibn3,bibx6}
In the latter case, it also agrees with the result from a WKB analysis.~\cite{bibx8}

\section{Kerr black holes}


Extending the above discussion to rotating (Kerr) black holes
is not straightforward.
Numerical results paint a complicated picture~\cite{bibx10} and analytical
calculations have yet to produce results.
We have obtained explicit results~\cite{bibus2} in the case $a = J/M \ll 1$, where
$J$ is the angular momentum of the black hole and $M$ is its mass.
This regime includes the Schwarzschild black hole ($a=0$).
For the asymptotic range of frequencies
\be\label{eqran} 1 \lesssim |\omega| \lesssim 1/a\ee
working as in the Schwarzschild case, we obtain~\cite{bibus2}
\be \Re\omega = {\textstyle\frac{1}{4\pi}}\ \ln (1+2\cos \pi j) + m\Omega + o(a^2) \ee
where $m$ is the azimuthal eigenvalue of the wave and $\Omega \approx a$ is the angular velocity of the horizon.
In the Schwarzschild limit, the range of frequencies~(\ref{eqran}) extends to infinity ($1/a\to\infty$) and we reproduce our earlier result~(\ref{eqqnm0}).
It would be interesting to extend this result to asymptotic frequencies for
arbitrary values of $a$ ($0<a<1$).

\section{AdS Black Holes}

According to the AdS/CFT correspondence,
QNMs for an AdS black hole
are expected to
correspond to perturbations of the dual CFT.
The establishment of such a
correspondence is hindered by difficulties in solving the wave equation.
In three dimensions, it reduces to a hypergeometric equation which is
analytically solvable~\cite{bibq7}.
Numerical results have been obtained in four, five and seven dimensions~\cite{bibq2}.

In five dimensions, the wave equation reduces to a Heun equation which cannot
be solved analytically.
We shall build a perturbative expansion in the case of a large black hole based on an
approximation to the wave equation which is valid in the
high frequency regime~\cite{bibus3}.
The singularities are located at $r^2 = \pm r_0^2$, where $r_0$ is the radius of the
horizon. In higher dimensions, there are more singularities, all lying on the
circle $|r| = r_0$ in the complex $r$-plane. Even though these are unphysical
singularities (with the exception of $r=r_0$), they seem to play an important
role in determining the QNMs, as in the Schwarzschild case~\cite{bibx3}.

Setting the AdS radius $R=1$, the wave equation for a massive scalar of mass $m$, frequency $\omega$ and transverse momentum $\vec p$ reads
\be
\left( y(y^2-1) \Psi' \right)' + \big\{ {\textstyle\frac{\hat\omega^2}{4}}\ {\textstyle\frac{y^2}{y^2-1}} - {\textstyle\frac{\hat p^2}{4}} - m^2 y\big\} \Psi = 0
\ee
where $y = \frac{r^2}{r_0^2}$, $\hat\omega = \frac{\omega}{\pi T_H}$,
$\hat p = \frac{|\vec p|}{\pi T_H}$.
After isolating the behavior at the two singularities, $y=\pm 1$,
\be
\Psi (y) = (y-1)^{-i\hat\omega/4} (y+1)^{-\hat\omega/4} F(y)
\ee
it reduces to the Heun equation
\bea
y(y^2-1) F'' + \left\{ \left( 3- {\textstyle\frac{i+ 1}{2}}\, \hat\omega \right) y^2 + {\textstyle\frac{1 -i}{2}}\, \hat\omega y -1 \right\} F'& & \nonumber\\
+ \left\{ {\textstyle\frac{\hat\omega}{2}}\left( {\textstyle\frac{i\hat\omega}{4}} - 1-i\right) y
-m^2y + (1-i){\textstyle\frac{\hat\omega}{4}} - {\textstyle\frac{\hat p^2}{4}} \right\}\; F &=& 0 
\eea
For large $\hat\omega$ and in the physical regime $r>r_0$, this may be
approximated by the hypergeometric equation~\cite{bibus3}
\be
(y^2-1) F_0'' + \left\{ \left( 3- {\textstyle\frac{i+ 1}{2}}\, \hat\omega \right) y + {\textstyle\frac{1 -i}{2}}\, \hat\omega \right\} F_0'
+ \left\{ {\textstyle\frac{\hat\omega}{2}} \left( {\textstyle\frac{i\hat\omega}{4}} - 1-i\right)
-m^2 \right\}\; F_0 = 0 
\ee
Two independent solutions are
\be\label{eqsolinf} \mathcal{K}_\pm = (x+1)^{-a_\pm} F(a_\pm, c-a_\mp ; a_\pm -a_\mp +1; 1/(x+1))\ee
where
$a_\pm = h_\pm - \frac{1+i}{4}\ \hat\omega$,
$c = \frac{3}{2} - \frac{i}{2}\ \hat\omega$, $h_\pm = 1\pm \sqrt{1+ m^2/4}$ and $x = \frac{y-1}{2}$.
The solution which is well-behaved at the boundary ($r\to\infty$) is $F_0 = \mathcal{K}_+$.
Near the horizon ($x\to 0$),
we have $F_0 \sim \mathcal{A}_0 + \mathcal{B}_0
x^{1-c}$,
where
\be\label{eq21} \mathcal{A}_0 = {\textstyle\frac{\Gamma(1-c)\Gamma(1-a_-+a_+)}{\Gamma(1-a_-)\Gamma(1-c+a_+)}}
\ \ , \ \ \mathcal{B}_0 = {\textstyle\frac{\Gamma(c-1)\Gamma(1+a_+-a_-)}{\Gamma(a_+)\Gamma(c-a_-)}}\ee
For a QNM, we demand regularity at the horizon, so $\mathcal{B}_0 = 0$.
This yields the spectrum
\be\label{eqo} \hat\omega_n = -2(1+ i)(n+h_+-{\textstyle{\frac{3}{2}}}) \ \ , \ \ n=1,2,\dots\ee
which agrees with numerical results~\cite{bibns}.

The first-order correction may be written as
\be\label{eq33} F_1(x) = \mathcal{K}_-(x) {\textstyle\int_x^\infty \frac{\mathcal{K}_+ \mathcal{H}_1 F_0}{\mathcal{W} }}
- \mathcal{K}_+(x) {\textstyle\int_x^\infty \frac{\mathcal{K}_-\mathcal{H}_1 F_0 }{\mathcal{W} }}
\ee
where $\mathcal{W}$ is the Wronskian and
\be\mathcal{H}_1 = {\textstyle\frac{1}{2x+1}} \big\{ {\textstyle\frac{1}{2} \frac{d}{dx}} + (i- 1){\textstyle\frac{\hat\omega}{4}} + {\textstyle\frac{\hat p^2}{4}}
\big\} \ee
At the horizon,
$F_1(x) \sim \mathcal{A}_1 + \mathcal{B}_1 x^{1-c}$, where
\be\label{eq39}
\mathcal{B}_1 = {\textstyle\frac{\Gamma(c-1)\Gamma(1+a_- -a_+)}{\Gamma(a_-)\Gamma(c-a_+)}}\ {\textstyle\int_0^\infty \frac{\mathcal{K}_+ \mathcal{H}_1 \mathcal{K}_+}{\mathcal{W}}}
\ee
The QNM frequencies to first order are solutions of
$\mathcal{B}_0 + \mathcal{B}_1 = 0$.
To solve this equation, we
need to calculate the integral in~(\ref{eq39}).
A somewhat tedious calculation~\cite{bibus3} leads to corrections to the
zeroth-order expression for QNM frequencies~(\ref{eqo}) which are of order $1/h_+\sim
1/m$.

\section{Conclusions}

Even though QNMs have long been known {\em numerically}, we have only
recently made progress toward calculating them {\em analytically}.
Their behavior appears to rely on
{\em unobservable}, or even
{\em unphysical}
singularities.
It is desirable to fully investigate the behavior of QNMs in general space-time backgrounds.
In asymptotically AdS spaces, we would like to understand their relevance to
the AdS/CFT correspondence.
In asymptotically flat spacetimes, they should lead to the equivalent of
Bohr's correspondence principle for the gravitational force which seems to
exist in the case of Schwarzschild black holes but remains
elusive for Kerr black holes. This may shed some light on the quantum
theory of gravity.

\end{document}